\documentclass[twocolumn,prb, showpacs]{revtex4}
\usepackage{psfrag}
\usepackage{graphicx}
\usepackage{amsmath}
\usepackage{amssymb}
\usepackage{eufrak}
\usepackage[ansinew]{inputenc}
\usepackage{natbib}

\begin{document}

\title{On the electron to proton mass ratio and the proton structure}

\author{Ole L. Trinhammer}
\affiliation{Department of Physics, Technical University of Denmark, \\
Fysikvej, Building 307, DK-2800 Kongens Lyngby, Denmark, EU\\} 

\begin{abstract}
We derive an expression for the electron to nucleon mass ratio from a reinterpreted lattice gauge theory Hamiltonian to describe interior baryon dynamics. We use the classical electron radius as our fundamental length scale. Based on expansions on trigonometric Slater determinants for a neutral state a specific numerical result is found to be less than three percent off the experimental value for the neutron. Via the exterior derivative on the Lie group configuration space u(3) we derive approximate parameter free parton distribution functions that compare rather well with those for the $u$ and $d$ valence quarks of the proton.
\end{abstract}

short title: Baryon phenomena I

\maketitle

\section*{1 The mass ratio and the model} 

The ratio we get between the electron mass $m_e$ and the proton mass $m_p$ is
\begin{equation} \label{eq:memp}
   \frac{m_e}{m_p} =\frac{\alpha}{\pi} \frac{1}{\rm{E}},
\end{equation}
where $\alpha=e^2/(4\pi\epsilon_0\hbar c)$ is the fine structure constant \citep{Sakurai} and $\rm{E} ={\it E}/\Lambda$ is the dimensionless ground state eigenvalue of a reinterpreted lattice gauge theory Kogut-Susskind Hamiltonian \citep{KogutSusskind1975}
\begin{equation} \label{eq:Hpsi}
   \frac{\hbar c}{a} \left[ - \frac{1}{2} \Delta + \frac{1}{2} \rm{Tr}\chi^2\right] \Psi(u) = E \Psi (u).
\end{equation}
with Manton's action \citep{Manton} used now as a potential for a configuration variable $u=e^{i\chi}$ in the Lie group u(3) in stead of a link variable $U$ in the SU(3) algebra \citep{Trinhammer1983}. The energy scale $\Lambda \equiv \hbar c/a$  corresponds to a fundamental length scale $a$, which we shall relate to the classical electron radius $R_0$ by
\begin{equation} \label{eq:a}
  a\pi=R_0,
\end{equation}
where $R_0$ is determined by the electron self potential energy \citep{LandauLifshitz}
\begin{equation} \label{eq:R0}
  \frac{1}{4\pi\epsilon_0}\frac{e^2}{R_0}=m_ec^2.
\end{equation}
We assume (\ref{eq:Hpsi}) to describe the baryon spectrum and identify the ground state with the proton. With $E=m_pc^2={\rm{E}}\Lambda={\rm{E}}\hbar c/a$ and   (\ref{eq:R0}) applied in (\ref{eq:a}), eq. (\ref{eq:memp}) follows directly. Our configuration space is "orthogonal" to the laboratory space wherefore (\ref{eq:Hpsi}) describes a truly interior dynamics which may be projected on laboratory space parameters through the eigenangles $\theta_j$ parametrizing the eigenvalues $e^{i\theta_j}$ of $u$. The projection introduces the dimensionful scale $a$, thus
\begin{equation} \label{eq:proj}
  x_j=a \theta_j
\end{equation}
\begin{figure} 
\begin{center}
\includegraphics[width=0.45\textwidth]{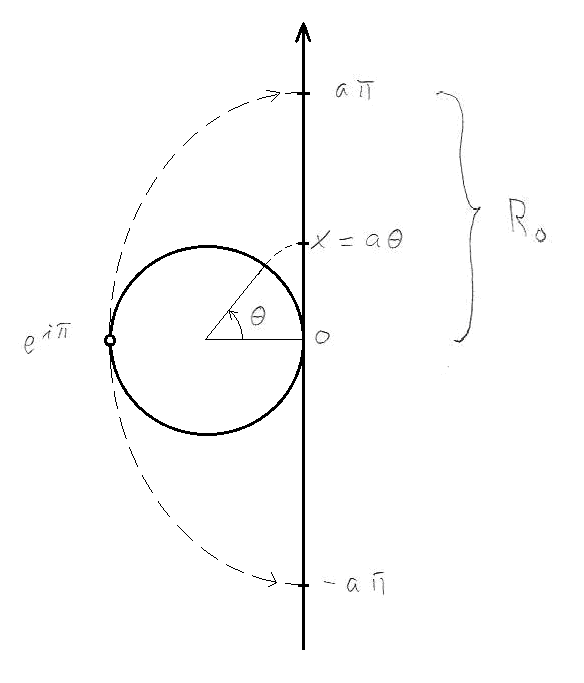}
\caption{Projection from configuration space to laboratory space.}
\label{fig:proj}
\end{center}
\end{figure}
Now a shortest geodesic \citep{Milnor} to track along the full extension of the u(3) maximal torus runs from the origin at $u=I$ where all eigenvalues are $1=e^{i0}$ to $u=-I$ where all eigenvalues are $-1=e^{i\pi}$, see also fig. \ref{fig:proj}. When for instance the neutron decays to the proton - and the electron is created as a "peel off" - the topological change in the interior baryon state maps by projection to laboratory space. It is not a new idea to suggest the classical electron radius as a fundamental length in elementary particle physics \citep{HeisenbergMehra, Heisenberg}. Here we specify the introduction of the scale (\ref{eq:a}) via the projection (\ref{eq:proj}). Conjugate to the space (angle) parameters in (\ref{eq:proj}) are canonical momentum (action) operators
\begin{equation}    
   p_j = - i \hbar \frac{1}{a} \frac{\partial}{\partial \theta_j} = \frac{\hbar}{a} T_j.
\end{equation}	
The toroidal generators $T_j$ induce coordinate fields $\partial_j$ according to the above mentioned eigenangle parametrization of the u(3) torus. In general the nine generators $T_k$ of u(3) namely induce coordinate fields as follows
\begin{equation}    
     \partial_k =  \frac{\partial}{\partial \theta} ue^{i\theta T_k} \vert _{\theta=0} = u i T_k.
\end{equation}
The remaining six off-toroidal generators are important in the baryon spectroscopy phenomenology resulting from (\ref{eq:Hpsi}) since they take care of spin and flavour degrees of freedom. With these parametrizations the  Laplacian ($\hbar=1$) in a polar decomposition reads \citep{Wadia, TrinhammerOlafsson}
\begin{equation}    
 \Delta= \sum^3_{j=1} \frac{1}{J^2} \frac{\partial}{\partial \theta_j} J^2 \frac{\partial}{\partial \theta_j} - \! \! \! \sum^3_{\substack{ i <  j \\ k\neq i,j}} \frac{K^2_k + M^2_k}{8 \sin^2 \frac{1}{2}(\theta_i -\theta_j)} 
\end{equation}
Here the Van de Monde determinant, the "Jacobian" of the parametrization is \cite{Weyl}
\begin{equation}    
      J = \prod^3_{i <  j } 2 \sin\left(\frac{1}{2} (\theta_i- \theta_j)\right) .
\end{equation}
The operators $K_k$ commute as body fixed angular momentum operators and $M_k$ "connect" the algebra by commuting into the subspace of $K_k$
\begin{equation}
  [M_k,M_l]=[K_k,K_l]=-iK_m,
\end{equation}
cyclic in $k,l,m$. The presence of the components of ${\bf{K}}=(K_1,K_2,K_3)$ and ${\bf{M}}=(M_1,M_2,M_3)$ in the Laplacian opens for the inclusion of spin and flavour. Interpreting $\bf{K}$ as the interior spin operator is encouraged by the body fixed signature of the commutation relations. The relation between laboratory space and interior space is like the relation in nuclear physics between fixed coordinate systems and intrinsic body fixed coordinate systems for the description of rotational degrees of freedom. To derive the spectrum for $M^2$ we use a coordinate representation \citep{Schiff} 
\begin{align} 
   K_1& = a \theta_2p_3 - a \theta_3 p_2 = \hbar \lambda_7 \nonumber \\ 
     K_2& = a \theta_1p_3 - a \theta_3 p_1 = \hbar \lambda_5 \nonumber \\
     K_3& = a \theta_1p_2 - a \theta_2 p_1 = \hbar \lambda_2. 
\end{align}
and
\begin{align} 
  M_3/\hbar &= \theta_1 \theta_2 +  \frac{a^2}{{\hbar}^2} p_1 p_2 = \lambda_1 \nonumber \\
  M_2/\hbar &= \theta_3 \theta_1 +  \frac{a^2}{{\hbar}^2} p_3 p_1 = \lambda_4 \nonumber \\
  M_1/\hbar &= \theta_2 \theta_3 +  \frac{a^2}{{\hbar}^2} p_2 p_3 = \lambda_6 .
\end{align}
The lambdas are corresponding Gell-Mann generators \citep{SchiffOpCit}. From these and
\begin{gather} 
  Y/\hbar =
     \frac{1}{6} ( \theta^2_1 + \theta^2_2- 2\theta^2_3) 
   + \frac{1}{6} \frac{a^2}{{\hbar}^2}(p^2_1 + p^2_2- 2p^2_3) =\lambda_8/\sqrt{3}, \nonumber \\
  2I_3/\hbar =  \frac{1}{2} ( \theta_1^2 - \theta^2_2) + \frac{1}{2}  \frac{a^2}{{\hbar}^2}
   (p_1^2 - p^2_2) = \lambda_3
\end{gather}
we find by straightforward but tideous commutations the spectrum
\begin{align} 
  M^2 = \frac{4}{3}(n+\frac{3}{2})^2 - K(K+1) - 3  -\frac{1}{3}y^2 - 4i^2_3, \nonumber \\
  n = 0,1,2,3, \dots \end{align}
where $y$ and $i_3$ are hypercharge and isospin three-component quantum numbers. The minimum value for the positive definite $M^2$ is 13/4 in the case of spin 1/2, hypercharge 1 and isospin 1/2 as for the nucleon. To solve for the eigenvalues we factorize the wavefunction
\begin{equation}
  \Psi(u)=\tau(\theta_1,\theta_2,\theta_3)\Upsilon(\alpha_4,\alpha_5,\alpha_6,\alpha_7,\alpha_8,\alpha_9),
\end{equation}
insert it in (2) and then integrate over the off-toroidal degrees of freedom $(\alpha_4,\alpha_5,\alpha_6,\alpha_7,\alpha_8,\alpha_9)$ to get for the measure scaled wavefunction $R=J\tau (\theta_1,\theta_2,\theta_3),$ for toroidal degrees of freedom
\begin{equation}
  \left[ -\sum^3_{j=1} \frac{\partial^2}{\partial \theta_j^2}+V\right]R=2{\rm{E}}R.
\end{equation}
Here  
\begin{gather}    
  V = -2 + \frac{1}{3}(K(K+1) + M^2) \sum^3_{ i <  j} \frac{1}{8 \sin^2 \frac{1}{2}(\theta_i -\theta_j)}
\nonumber \\\hspace{1cm} +  2 (v(\theta_1) + v(\theta_2)+ v(\theta_3)). \hspace{2.5cm}
\end{gather}
\begin{figure}
\begin{center}
\includegraphics[width=0.45\textwidth]{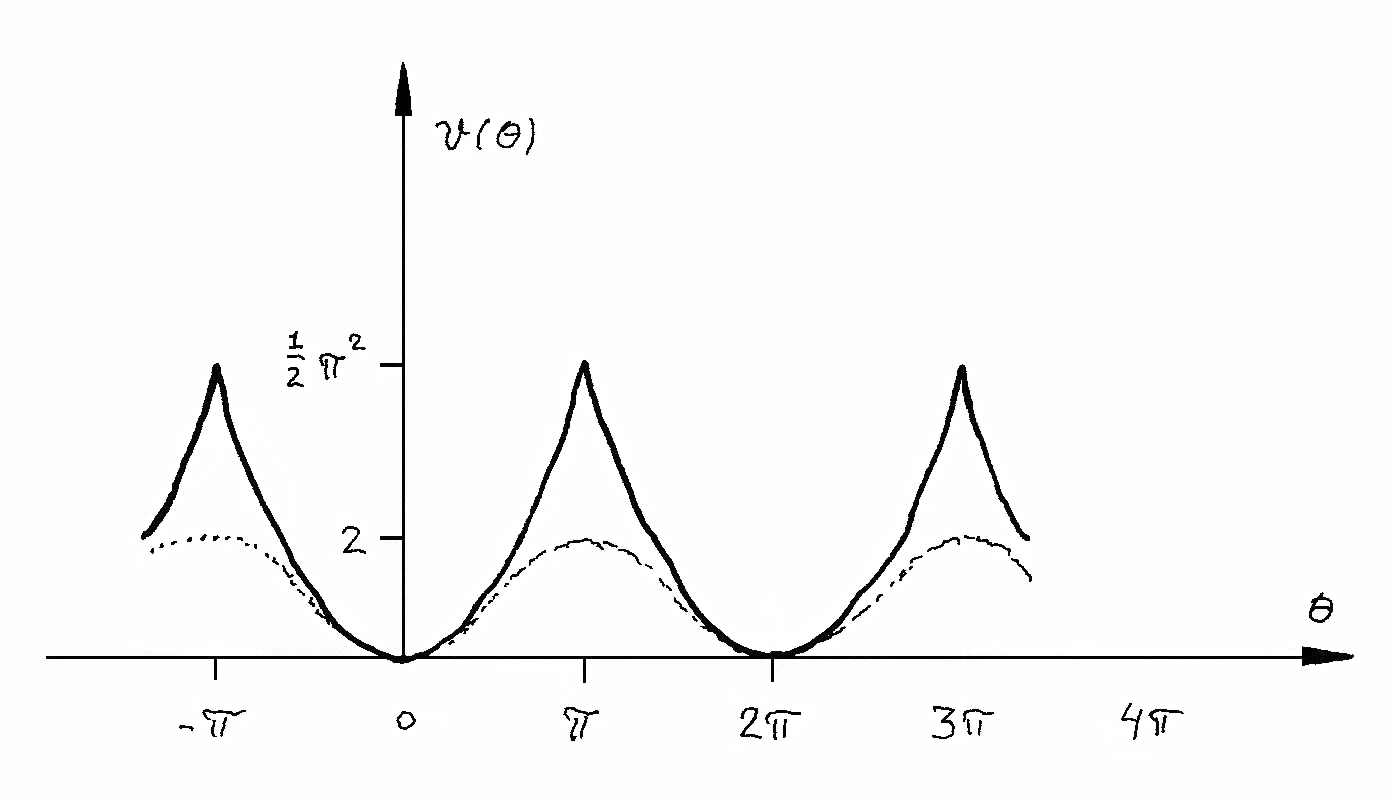}
\caption{Periodic parametric potential (\ref{eq:parapotential}) originating from the potential in (\ref{eq:Hpsi}). The dashed curve corresponds to the Wilson analogue \citep{Wilson} of the Manton action and is not considered in the present work.}
\label{fig:parapotential}
\end{center}
\end{figure}
where (see fig. \ref{fig:parapotential})
\begin{align}   \label{eq:parapotential}  
   v(\theta)     = \frac{1}{2}(\theta - n\cdot 2\pi)^2, \quad \theta \in [(2n-1)\pi,(2n+1)\pi], \nonumber \\
   {\rm{for}}\quad n\in Z. \hspace{4cm}
\end{align}
By expansion on Slater determinants \cite{Wadia}
\begin{equation} \label{eq:slater}
  b_{pqr}=\epsilon_{ijk}\cos(p\theta_i)\sin(q\theta_j)\cos(r\theta_k)
\end{equation}
with integer $p,q,r$, we can solve (16) by a Rayleigh Ritz method \cite{BruunNielsen} to yield the ground state eigenvalue ${\rm{E_0}}=4.38$ which corresponds to $m_e/m_0=1/1885$. This is less than three percent off the value $m_e/m_n=1/1838.6...$ based on experimental data for the electron and neutron \cite{Beringer}. Note that $b_{pqr}$ is antisymmetric in the colour degrees of freedom $\theta_j$.

\section*{2 Parton distribution functions}

\begin{figure}
\begin{center}
\includegraphics[width=0.45\textwidth]{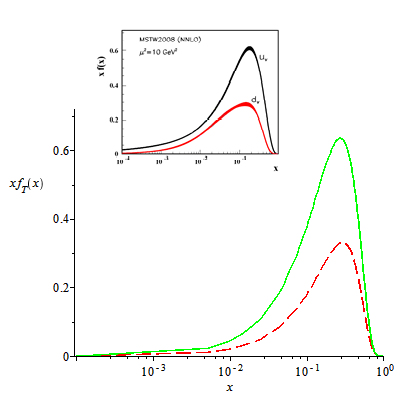}
\caption{Parton distribution functions from a first order approximation to the ground state of (\ref{eq:Hpsi}) as a function of the parton fraction $x$. The distributions are generated by impact along $T_u=(2/3,0,-1)$ (solid, green line) and $T_d=(-1/3,0,-1)$ (dashed, red line) with no fitting parameters. For comparison we show $u$ and $d$ valence quarks of the proton at 10 GeV$^2$ in the insert adapted from the particle data group with other parton distributions erased \cite{Beringer}.}
\label{fig:pdf}
\end{center}
\end{figure}

We can generate parton distribution functions by projections via the momentum form, i.e. the exterior derivative on the u(3) manifold. For this we expand the exterior derivative $dR$ of the measure scaled toroidal wavefunction $R$ on the torus forms $d\theta_j$,
\begin{equation}
  dR=\psi_jd\theta_j.
\end{equation}
The action of the torus forms on the toroidal coordinate fields expresses the generalization to the interior configuration space manifold of the quantization inherent in the conjugate variables in (5) and (6), thus
\begin{equation}
  d\theta_i(\partial_j)=\delta_{ij} \Leftrightarrow [\partial_j,\theta_i]=\delta_{ij}
\end{equation}
Inspired by Bettini's elegant treatment of parton scattering \cite{Bettini} we generate distribution functions via our exterior derivative. The derivation runs like this (with $\hbar = c=1$): Imagine a proton at rest with four-momentum 
$P = ({\text{\bf{0}}}, E_0)$. We boost it virtually to energy $E$ by impacting upon it a massless four-momentum $q = ({\text{\bf{q}}}, E - E_0) $ which we assume to hit a parton $xP$. After impact the parton represents a virtual mass $xE$. Thus
\begin{equation}
    (xP_{\mu} + q_{\mu})\cdot(xP^{\mu} + q^{\mu}) = x^2E^2,
\end{equation}
from which we get the parton momentum fraction
\begin{equation}
   x = \frac{2E_0}{E + E_0},
\end{equation}
or the boost parameter
\begin{equation}
    \xi\equiv \frac{E - E_0}{E} = \frac{2 - 2x}{2-x}.
\end{equation}
\begin{figure} [h]
\begin{center}
\includegraphics[width=0.35\textwidth]{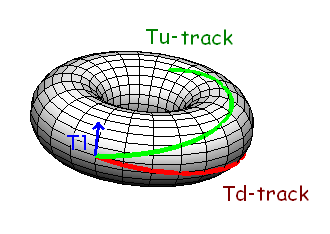}
\caption{The generators $T_u$ and $T_d$ given in (\ref{eq:TuTd}) generate traces along the u(3) torus as shown here in the first and third toroidal degrees of freedom with $T_3$ pointing to the left.}
\label{fig:2Dtorus}
\end{center}
\end{figure} 
We can use the boost parameter for an angular track $\theta=\pi\xi$ on the manifold in the direction laid out by a specific toroidal generator
\begin{equation}
  T=a_1T_1+a_2T_2+a_3T_3.
\end{equation}
With the toroidal generator $T$ as introtangling momentum operator we namely have the qualitative correspondence $q_{\mu} \sim E - E_0 \sim (1-x)E \sim(1-x)T$. That is, we will project along $\xi T \sim(1-x)T$ in order to probe on $xP_{\mu}$. 
With a probability amplitude interpretation of $R$ we project on a fixed colour base $iT_j$ and sum over the colour components for a specific generator $T$ to get the corresponding distribution function $f_T(x)$ determined by 
\begin{equation}
   f_T(x)dx=(\sum^3_{j=1} dR_{u=exp(\theta iT)}(iT_j))^2 d\theta
\end{equation}
By a pull-back operation \cite{GuilleminPollack} to parameter space we get
\begin{gather}
   \sum^3_{j=1}dR_{u = \exp(\theta i T)}(iT_j)=dR_{u = \exp(\theta i T)}(i(T_1+T_2+T_3)) \nonumber\\ 
      =\frac{d}{dt}R(ue^{ti(T_1+T_2+T_3)}) {\text{\Large{$\mid$}}}_{t=0} \nonumber \\
   =\frac{d}{dt}R(a_1\theta +t, a_2\theta +t, a_3\theta +t) {\text{\Large{$\mid$}}}_{t=0} \nonumber \\
   =\sum^3_{j=1}\frac{\partial R}{\partial \theta_j}{\text{\Large{$\mid$}}}_{(\theta_1, \theta_2, \theta_3)
  =(\theta \cdot a_1, \theta \cdot a_2, \theta \cdot a_3)}\cdot \frac{\partial (a_j\theta +t)}{\partial t}{\text{\Large{$\mid$}}}_{t=0} \nonumber \\
 =\left(  \frac{\partial R}{\partial \theta_1}+ \frac{\partial R}{\partial \theta_2}+ \frac{\partial R}{\partial                    \theta_3} \right){\text{\Large{$\mid$}}}_{(\theta_1, \theta_2, \theta_3)
  =(\theta \cdot a_1, \theta \cdot a_2, \theta \cdot a_3)}     \nonumber \\
 \equiv D(\theta \cdot a_1, \theta \cdot a_2, \theta \cdot a_3).
\end{gather}
Along the tracks shown in fig. \ref{fig:2Dtorus}, we generate distributions by
\begin{equation} \label{eq:TuTd}
   T_u = \begin{Bmatrix} 2/3 & 0 & 0 \\ 0 & 0 &0 \\0& 0& -1 \end{Bmatrix} \hspace{2mm} \text{and} \hspace{2mm}
     T_d = \begin{Bmatrix} -1/3 & 0 & 0 \\ 0 & 0 &0 \\0& 0& -1 \end{Bmatrix}\!.
\end{equation}
as shown in fig. \ref{fig:pdf} for a first order approximation
\begin{equation} \label{eq:protonic}
  b_{0, \frac{1}{2},1}  (\theta_1, \theta_2, \theta_3)  = \frac{1}{N} \begin{vmatrix}  1&1 & 1\\
\sin \frac{1}{2} \theta_1 & \sin \frac{1}{2} \theta_2 & \sin \frac{1}{2} \theta_3 \\
 \cos \theta_1 & \cos \theta_2 & \cos \theta_3\end{vmatrix}\!,
\end{equation}
to an expansion for a protonic state with normalization constant $N$. For instance
\begin{widetext}
\begin{equation}
  xf_{Tu}(x)= 
x\left[D\left( \pi \frac{2 - 2x}{2-x} \cdot \frac{2}{3}, \pi  \frac{2 - 2x}{2-x} \cdot 0, 
   \pi \frac{2 - 2x}{2-x} \cdot (-1) \right) \right]^2 \cdot \frac{\pi \cdot 2}{(2-x)^2}.
\end{equation}
\end{widetext}
where in its full glory
\begin{align}
ND(\theta_1, \theta_2, \theta_3) = \hspace{6cm} \nonumber \\
- \frac{1}{2} \cos \frac{\theta_1}{2}\cdot (\cos \theta_3 - \cos \theta_2) - \sin \theta_1 \cdot (\sin \frac{\theta_3}{2}-\sin \frac{\theta_2}{2} ) \nonumber \\
+  \frac{1}{2} \cos \frac{\theta_2}{2}\cdot (\cos \theta_3 - \cos \theta_1)+  \sin \theta_2 \cdot (\sin \frac{\theta_3}{2}-\sin \frac{\theta_1}{2} )\nonumber \\
-  \frac{1}{2} \cos \frac{\theta_3}{2}\cdot (\cos \theta_2 - \cos \theta_1) - \sin \theta_3 \cdot (\sin \frac{\theta_2}{2}-\sin \frac{\theta_1}{2} ).   
\end{align}
The distribution functions shown in fig.\ref{fig:pdf} contain no fitting parameters at all. Note that the $T_u$-parton momentum content of the approximate state (\ref{eq:protonic}) is close to the double of the $T_d$-content
\begin{gather}
   \int^1_0xf_{Tu}(x) dx \, {\text{\Large{/}}} \int^1_0xf_{Td}(x) dx \nonumber \\ = 0.2722 {\text{\Large{/}}} 0.1437  = 
 1.89 \approx 2 ,\end{gather}

 Work is in progress to solve (\ref{eq:Hpsi}) for the charged proton by antisymmetrized parity definite expansions on Bloch states \cite{Jacobsen}
\begin{equation}
  g_{pqr} = e^{i\mbox{\boldmath{$\kappa$}}\cdot\mbox{\boldmath{$\theta$}}}e^{ip\theta_1}
  e^{iq\theta_2}e^{ir\theta_3},
\end{equation}
where \mbox{\boldmath{$\kappa$}} $=(\kappa_1, \kappa_2, \kappa_3)$ are appropriately chosen Bloch vectors and \mbox{\boldmath{$\theta$}} $=(\theta_1, \theta_2, \theta_3)$ .

\section*{3 Comments}

It is not clear whether the minor but still significant discrepancy in our value for $m_e/m_n$ is due to disparate partial wave projections of the base set (\ref{eq:slater}) or due to corrections needed in (\ref{eq:a}). It is somewhat surprising that when applied to a neutral state expression (\ref{eq:memp}) gives a result just three percent off the experimental value for the electron to neutron value. This is surprising since the classical electron radius on which we base our prediction is normally presented to be just an order of magnitude scale for strong interaction phenomena \cite{HeisenbergOpCit}. It should be noted though that in (\ref{eq:proj}) we have introduced a well defined projection to space parameters. However, at the present level of understanding it seems wise to quote L. D. Landau and E. M. Lifshitz: "...it is impossible within the framework of classical electrodynamics to pose the question whether the total mass of the electron is electrodynamic." \cite{LandauLifshitz}

\section*{Conclusion}

We have derived an expression for the electron to nucleon mass ratio from a reinterpreted lattice gauge theory Hamiltonian. A specific calculation for the neutron from expansions on Slater determinants of indefinite parity gives a result less than three percent off the experimental value. The proximity of prediction to observation should encourage further study within the framework of the Hamiltonian model presented. From the same model we have derived approximate parameter free parton distribution functions that compare rather well with those for the $u$ and $d$ valence quarks of the proton. Work is in progress to establish a complete Bloch wave base for expansions with suitable symmetry requirements implied by the potential and the antisymmetry under interchange of colour degrees of freedom.

\section*{Acknowledgment}
I am grateful to the late Victor F. Weisskopf for inspiration on $m_e/m_p$ and to Holger Bech Nielsen for clarifying discussions on the momentum form. I thank Torben Amtrup for moral support through the years, Jakob Bohr for advice and Erik Both for help in preparing the manuscript for publication.

\section*{References}
\bibliographystyle{plain}

\end{document}